%% file: 2603-029-3H-HibikiHayasaki.tex
\documentclass[pteplogo]{ptephy_v2}%%%%%% to generate preprint number with ptep logo
\usepackage{hyperref}
\usepackage{graphicx}
\usepackage{subcaption}
\usepackage[version=4]{mhchem}
\begin{document}

\title{Performance evaluation of GYGAG crystals for neutrinoless double beta decay search}

\author[*1]{H. Hayasaki}
\author[2]{K. Hosokawa}
\author[2]{K. Ichimura} 
\author[3]{S. Kurosawa} 
\author[1]{Y. Masaki}
\author[1]{Y. Mizuno}
\author[1]{Y. Nakajima}
\author[4]{H. Sekiya} 
\author[2]{A. Yamaji}

\affil{Department of Physics, Graduate School of Science, The University of Tokyo, 7-3-1 Hongo, Bunkyo-ku, Tokyo, Japan\email{hayasaki@hep.phys.s.u-tokyo.ac.jp}}
\affil[2]{Research Center for Neutrino Science, Tohoku University}
\affil[3]{Faculty of Engineering, The University of Tokyo}
\affil[4]{Institute for Cosmic Ray Research, The University of Tokyo}

\begin{abstract}
Neutrinoless double beta decay (0$\nu\beta\beta$) is a crucial for confirming neutrinos' Majorana characteristics.
Previous searches for 0$\nu\beta\beta$ decay of $^{160}$Gd using GSO crystals was conducted, but the sensitivity was limited by background $\alpha$-rays from radioactive impurities.
To distinguish the backgrounds, PSD (Pulse Shape Discrimination) and energy resolution are essential.
GAGG (Gd$_{3}$(Al, Ga)$_5$O$_{12}$) is a $^{160}$Gd-containing crystal that exhibits a higher light yield than GSO and excellent PSD capabilities.
GYGAG (Gd$_{3-x}$Y$_x$(Al, Ga)$_5$O$_{12}$) is a variant crystal where a portion of Gd is substituted with Yttrium and can have enhanced light yield, motivating its investigation for 0$\nu\beta\beta$ decay searches.
In this study, GYGAG crystals with different Yttrium concentrations were fabricated and their energy resolution, pulse shape discrimination capability between $\alpha$ and $\gamma$ rays, and scintillation timing characteristics were evaluated for applications in 0$\nu\beta\beta$ decay searches.
Y substitution enhanced the light yield, and the best energy resolution was achieved for Gd$_{2.5}$Y$_{0.5}$Al$_2$Ga$_3$O$_{12}$.
GYGAG exhibited high pulse shape discrimination performance across various yttrium concentrations.
\end{abstract}
\subjectindex{H20}

\maketitle

\input{chapters/chapter1.tex}
\input{chapters/chapter2.tex}

\input{chapters/chapter3.tex}

\input{chapters/chapter4.tex}
\input{chapters/chapter5.tex}
\input{chapters/chapter6.tex}

\input{chapters/bib.tex}
\end{document}

%% file: chapters/chapter1.tex
\section{Introduction}
Searching for neutrinoless double beta decay (0$\nu\beta\beta$) is one of the important subjects in particle physics.
If observed, it would prove that neutrinos are Majorana particles and provide information on the Majorana mass of neutrinos.
Experiments using $^{136}$Xe are leading the world in the search for 0$\nu\beta\beta$, the upper limit on the Majorana mass is given as 28--122 meV from the results of KamLAND-Zen.\cite{KamLAND}
However, due to the large uncertainty in the nuclear matrix elements, it is important to conduct searches using multiple isotopes.

$^{160}$Gd is a nucleus that may undergo $0\nu\beta\beta$, with a relatively high natural abundance of 21.9\% and a Q-value of 1.73~MeV.
The current lower limit on the half-life of the neutrinoless double beta decay of $^{160}$Gd is reported to be $T_{1/2} > 1.3 \times 10^{21}$ years in an experiment using a 95 cm$^3$ GSO crystal~\cite{Danevich2001}.
The sensitivity of neutrinoless double-beta ($0\nu\beta\beta$) decay searches is primarily limited by $\alpha$-ray backgrounds from internal radioactive impurities. To mitigate these backgrounds, superior energy resolution and Pulse Shape Discrimination (PSD) performance are essential.
Energy resolution is crucial for distinguishing the $0\nu\beta\beta$ signal peak from the $2\nu\beta\beta$ continuum.
High light yield is fundamental to achieving these goals; it reduces statistical fluctuations in the number of detected photons, which directly improves energy resolution.
Ce-doped GAGG (Gd$_3$(Al, Ga)$_5$O$_{12}$) is a promising candidate among $^{160}$Gd-containing crystals due to its superior light yield ($\sim$57,000 photons/MeV) compared to GSO ($\sim$10,000 photons/MeV) and its excellent pulse shape discrimination capabilities.\cite{GAGG_alpha1}\cite{GAGG_alpha2} In fact, 0$\nu\beta\beta$ search using GAGG crystals is underway at the Kamioka Underground Laboratory~\cite{Kamioka}.

GYGAG (Gd$_{3-x}$Y$_x$(Al, Ga)$_5$O$_{12}$) is a variant crystal where a portion of Gd is substituted with Y, and reports indicate that this substitution leads to enhanced light yield, making it potentially well-suited for 0$\nu\beta\beta$ decay searches~\cite{GYGAG_1}.

The aim of this work is to find the best yttrium concentration for the search of 0$\nu\beta\beta$ to improve the sensitivity of future searches.
In this work, GYGAG crystals with different Y concentrations were fabricated, and their performance was evaluated for 0$\nu\beta\beta$ decay search applications, focusing on energy resolution, pulse shape discrimination capability, and scintillation timing characteristics.

%% file: chapters/chapter2.tex
\section{GYGAG Crystal}
\subsection{GAGG}
GAGG is a non-hygroscopic inorganic scintillator with optical properties, including a high light yield of approximately 57,000 photons/MeV, a peak emission wavelength of about 520 nm, and a density of 6.63 g/cm$^3$ \cite{GAGG_basic1}\cite{GAGG_basic2}.
The high density and effective atomic number make GAGG well-suited for gamma-ray detection.

One of the most remarkable properties of GAGG for 0$\nu\beta\beta$ search is its ability to discriminate between $\alpha$ rays and $\gamma$ rays events through pulse shape analysis at room temperature\cite{GAGG_alpha3}, which is crucial for background rejection in rare event searches.
\subsection{GYGAG}
GYGAG (Gd$_{3-x}$Y$_x$(Al, Ga)$_5$O$_{12}$) is formed by partially substituting Gd$^{3+}$ ions with Y$^{3+}$ ions in the dodecahedral sites of the garnet structure.
Scintillation properties of GYGAG single crystal were reported by several groups and it is reported that GYGAG has higher light yield compared to GAGG.
For example, a single crystal of Gd$_{1.5}$Y$_{1.5}$Al$_5$O$_{12}$(Ce) yttrium substitution reaches 105,000 photons/MeV~\cite{GYGAG_add1}.
Scintillation properties of ceramic samples were also reported by Hull et al.~\cite{GYGAG_add2}.
\subsection{Crystal Growth and Sample Preparation}
In this study, single crystals of Gd$_{3-x}$Y$_x$Al$_2$Ga$_3$O$_{12}$ were grown by the Czochralski process~\cite{Czochralski}\cite{Kurosawa2014}
from 99.99$\%$ pure starting materials, $\alpha$-Al$_2$O$_3$, $\beta$-Ga$_2$O$_3$, Gd$_2$O$_3$, and CeO$_2$, using an Ir-crucible, where $x = 0.5, 1.0, 1.5$. Here, a mixture gas of Ar~(98$\%$)+O$_2$~(2$\%$) was used as an atmosphere during crystal growth. After the growth, the samples were cut and polished with a size of 5 mm $\times$ 5 mm $\times$ 5 mm cubes.
The crystals are hereafter designated as Y0.0, Y0.5, Y1.0, and Y1.5 based on their yttrium concentrations for clarity.
Two cubes are made for Y0.5, Y1.0, and Y1.5. The crystals are hereafter designated as (Y0.5, 1) and (Y0.5, 2).
\subsection{Composition Analysis}
Quantitative compositional analysis was performed using a field-emission electron probe microanalyzer (FE-EPMA). 
The following compound standards were used for calibration: YAG for Al and Y, GaP for Ga, \ce{CePO4} for Ce, and \ce{GdPO4} for Gd.
The interference of the Ce $L\gamma$ line on the Gd $L\alpha$ line was corrected, followed by the application of the ZAF method~\cite{ZAF}.

\begin{figure}[ht]
    \centering
    \includegraphics[width=0.5\linewidth]{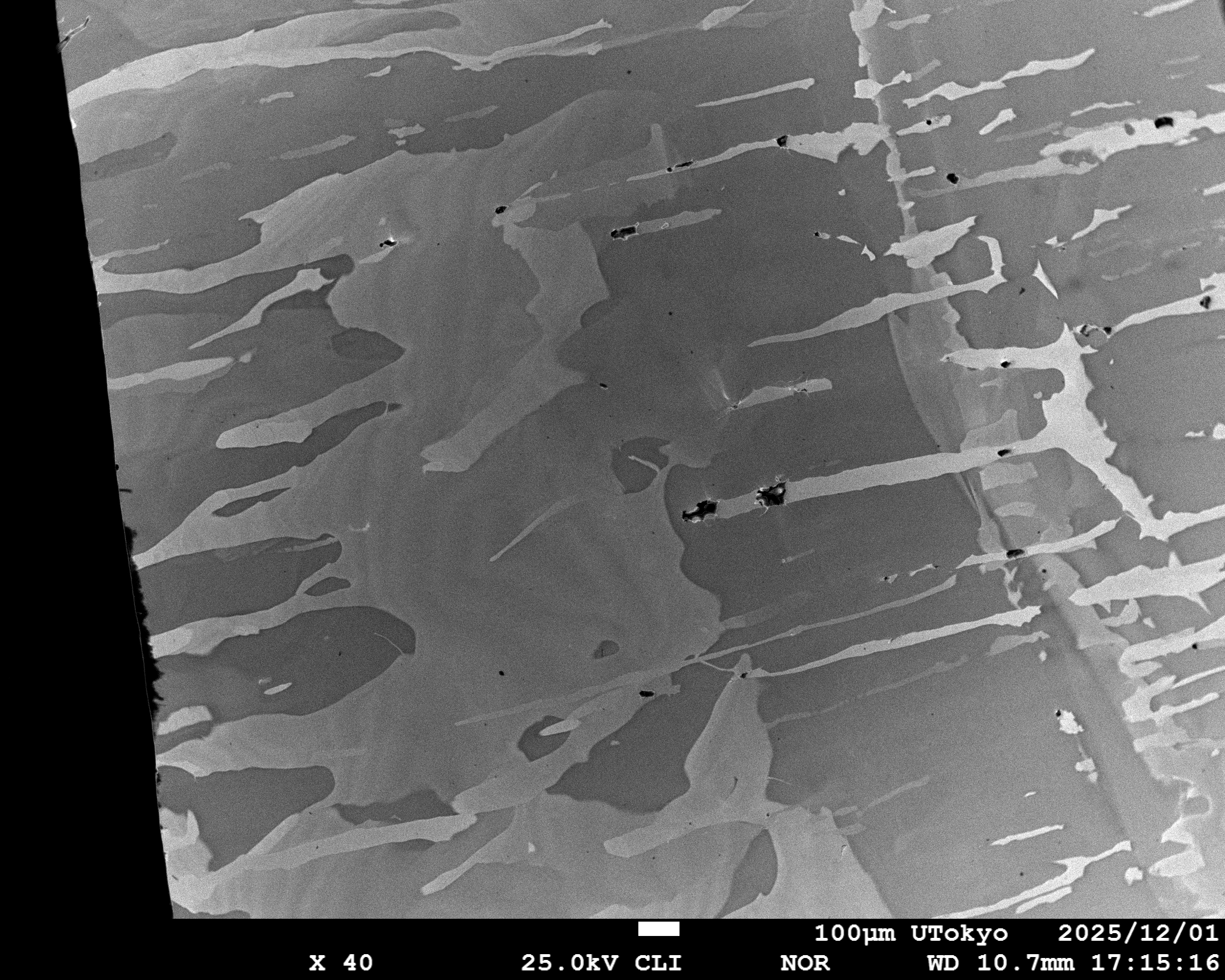}
    \caption{Cathodoluminescence image of the crystal. Bright and dark regions are observed, indicating inhomogeneity in the crystal composition.}
    \label{fig:cat}
\end{figure}

In addition to the FE-EPMA measurements, backscattered electron (BSE) and cathodoluminescence (CL) imaging were performed. While no significant contrast was observed in the BSE and CL images of the Y0.0 and Y0.5 samples, distinct brightness contrast appeared in the Y1.0 and Y1.5 samples, as shown in Figure~\ref{fig:cat}. This suggests that chemical segregation likely occurred during the crystal growth of the Y1.0 and Y1.5 samples.

For each crystal, EPMA measurements were performed and BSE and CL images are taken at twenty points.
The twenty points were selected at equal intervals along the diagonal of the square.
The composition of each crystal is calculated from the average of the points.
Table~\ref{tab:composition} summarizes the average compositions.
For the Y1.0 and Y1.5 samples, measurements were separately evaluated for the bright and dark regions.

The Y0.0 sample did not contain detectable yttrium.
In contrast, the Y0.5 sample exhibited a higher concentration of yttrium than the nominal value introduced during crystal preparation.
For the Y1.0 sample, the yttrium concentration was also higher than the nominal composition and was comparable to that of the Y1.5 sample, resulting in no significant difference between the two.

\begin{table}[ht]
    \centering
    \caption{Crystal compositions measured by FE-EPMA.
    Each value represents the number of atoms per formula unit of the garnet structure (normalized to $\text{O}=12$).
    Data are averaged over multiple measurement points, and uncertainties represent the standard deviation.
    }
    \begin{tabular}{|c|ccccc|}\hline
    Crystal & Gd & Y & Ga & Al & Ce  \\\hline
    Y0.0 & 3.098 $\pm$ 0.013 & 0.000 $\pm$ 0.000 & 2.221 $\pm$ 0.013 & 2.673 $\pm$ 0.016 & 0.008 $\pm$ 0.001 \\
Y0.5,1 & 2.407 $\pm$ 0.016 & 0.652 $\pm$ 0.018 & 2.921 $\pm$ 0.023 & 2.014 $\pm$ 0.030 & 0.006 $\pm$ 0.001 \\
Y0.5,2 & 2.382 $\pm$ 0.021 & 0.672 $\pm$ 0.012 & 2.901 $\pm$ 0.018 & 2.039 $\pm$ 0.036 & 0.006 $\pm$ 0.001 \\
Y1.0,1 bright & 1.573 $\pm$ 0.009 & 1.540 $\pm$ 0.021 & 2.887 $\pm$ 0.015 & 1.990 $\pm$ 0.019 & 0.011 $\pm$ 0.001 \\
Y1.0,1 dark & 1.517 $\pm$ 0.015 & 1.550 $\pm$ 0.016 & 2.898 $\pm$ 0.016 & 2.030 $\pm$ 0.029 & 0.006 $\pm$ 0.001 \\
Y1.0,2 bright & 1.520 $\pm$ 0.029 & 1.547 $\pm$ 0.057 & 2.905 $\pm$ 0.028 & 2.022 $\pm$ 0.016 & 0.006 $\pm$ 0.001 \\
Y1.0,2 dark & 1.504 $\pm$ 0.010 & 1.557 $\pm$ 0.014 & 2.875 $\pm$ 0.013 & 2.060 $\pm$ 0.016 & 0.005 $\pm$ 0.001 \\
Y1.5,1 bright & 1.521 $\pm$ 0.024 & 1.502 $\pm$ 0.036 & 2.968 $\pm$ 0.029 & 1.997 $\pm$ 0.023 & 0.012 $\pm$ 0.003 \\
Y1.5,1 dark & 1.471 $\pm$ 0.014 & 1.568 $\pm$ 0.023 & 2.906 $\pm$ 0.015 & 2.048 $\pm$ 0.014 & 0.008 $\pm$ 0.001 \\
Y1.5,2 bright & 1.490 $\pm$ 0.027 & 1.520 $\pm$ 0.019 & 2.919 $\pm$ 0.041 & 2.059 $\pm$ 0.056 & 0.012 $\pm$ 0.004 \\
Y1.5,2 dark & 1.462 $\pm$ 0.020 & 1.559 $\pm$ 0.033 & 2.887 $\pm$ 0.025 & 2.084 $\pm$ 0.025 & 0.008 $\pm$ 0.002 \\ \hline
    \end{tabular}
    \label{tab:composition}
\end{table}

%% file: chapters/chapter3.tex
\section{Evaluation of Energy Resolution}
\subsection{Method}
\begin{figure}[ht]
    \centering
    \begin{subfigure}[b]{0.49\textwidth}
        \includegraphics[width=\textwidth]{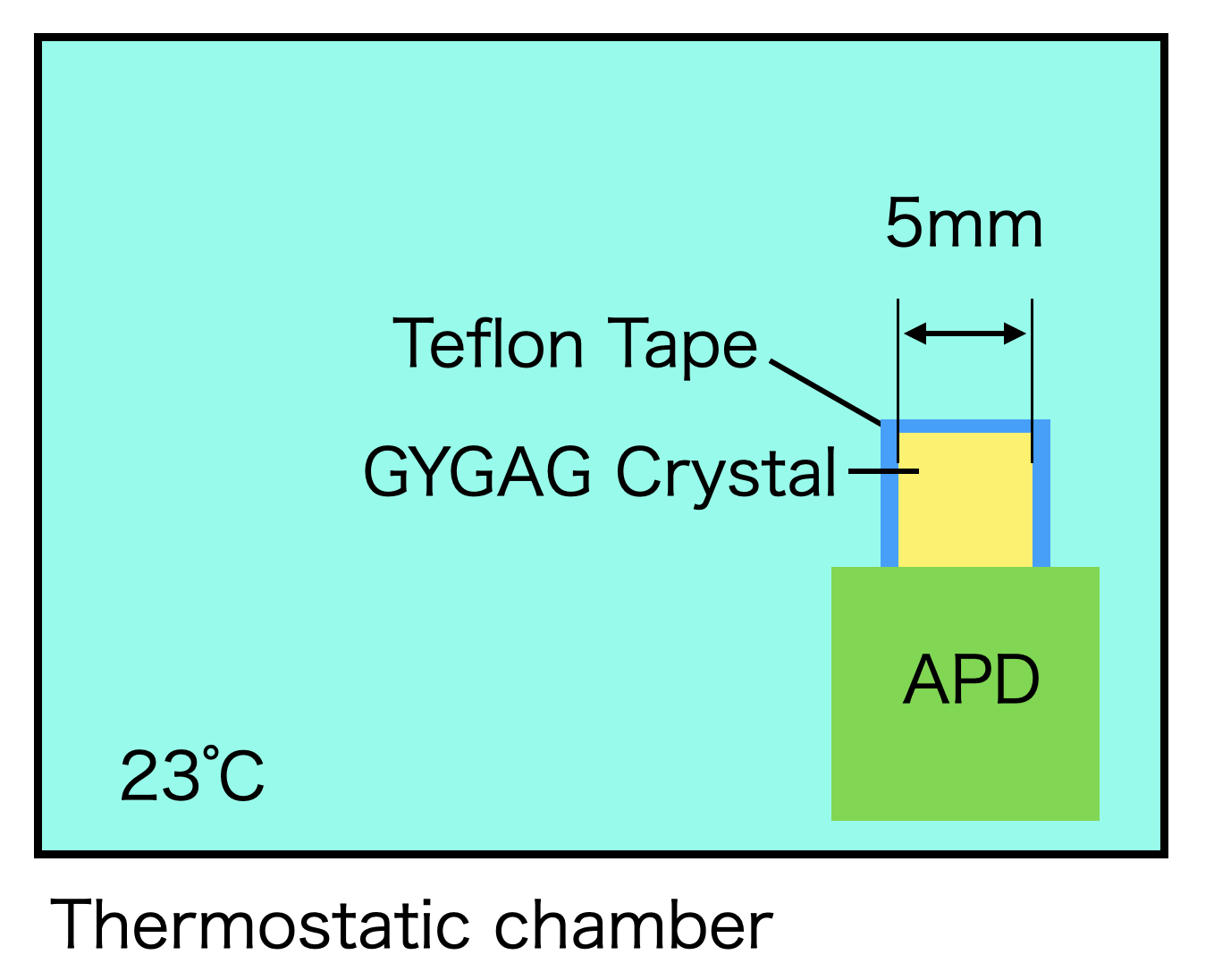}
        \caption{Schematic view of the measurement setup. 
        %The APD and the crystal are optically coupled using optical grease. The entire setup is placed inside a temperature-controlled chamber.
        }
        \label{fig:APD_setup}
    \end{subfigure}
    \hfill
    \begin{subfigure}[b]{0.49\textwidth}
        \includegraphics[width=\textwidth]{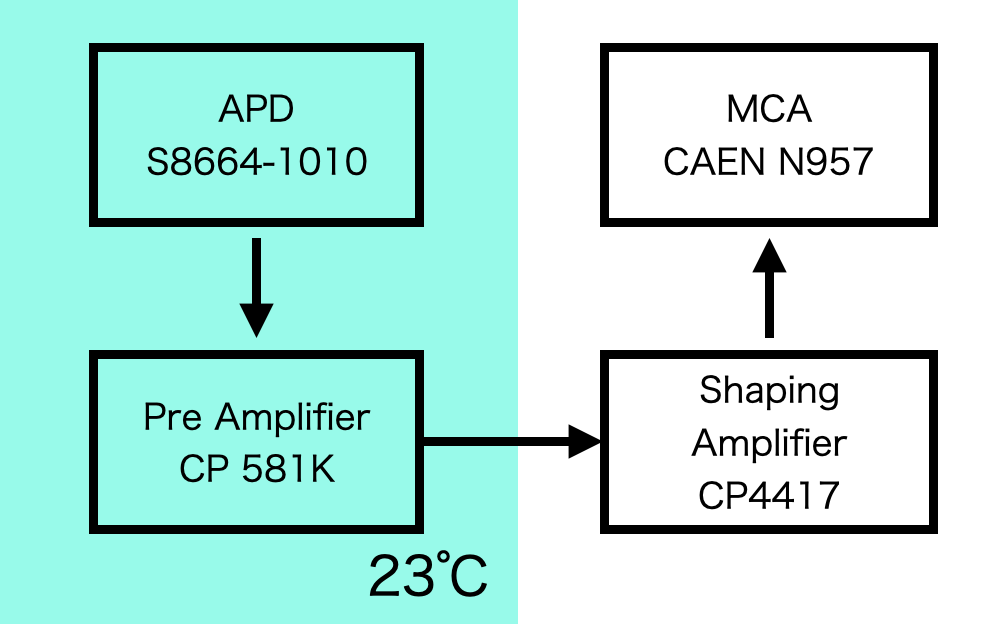}
        \caption{Processing of the signal
        %: the time constant of the shaping amplifier was set to 1 $\mu$s and the gain was set to 1.
        }
        \label{process}
    \end{subfigure}
    \caption{(a) The APD and the crystal are optically coupled using optical grease. The entire setup is placed inside a temperature-controlled chamber. (b) Signal processing chain.}
    \label{fig:APD_all}
\end{figure}

The crystals were irradiated with $^{137}$Cs and $^{22}$Na sources, and resulting scintillation signals were detected using an avalanche photodiode (APD; Hamamatsu S8664-1010).
GAGG emits scintillation light in the wavelength range of 500 -- 700 nm, and the APD used has high sensitivity in this range.
The crystals were wrapped with at least seven layers of 100 $\mu$m-thick Teflon tape. They were mounted on the APD without optical grease, as optical grease was found to cause temporal variations in the light collection efficiency.
The APD and preamplifier were housed in a temperature-controlled chamber to maintain a constant temperature of (23.0 $\pm$ 0.5)°C.
Figure~\ref{fig:APD_setup} shows a schematic view of the setup.

The signal processing method is shown in Figure \ref{process}.
The APD output was amplified by a preamplifier (Clear Pulse 581K) and shaped by a shaping amplifier (Clear Pulse 4417). 
The time constant of the shaping amplifier was set to 1 $\mu$s and the gain was set to 1.
The output from the shaping amplifier was digitized using a MCA (CAEN N957).
Crystals were measured at different bias voltages.
The energy resolution was found to improve with increasing bias voltage up to 315 V, after which it remained constant.
Therefore, the crystals were measured at a bias voltage of 315 V.

\subsection{Analysis Procedure and Results}
\begin{figure}[ht]
\centering
\begin{subfigure}[b]{0.48\textwidth}
    \centering
    \includegraphics[width=\textwidth]{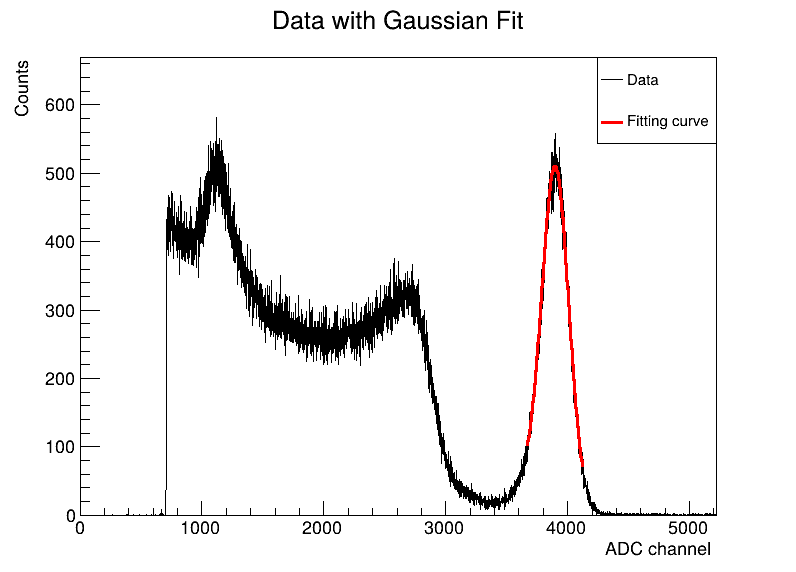}
    \caption{Example of fitting
    %: The black line is the data, the red line is the fitting function. Energy spectra of $^{137}$Cs.
    }
    \label{Fitting_example}
\end{subfigure}
\hfill
\begin{subfigure}[b]{0.48\textwidth}
    \centering
    \includegraphics[width=\textwidth]{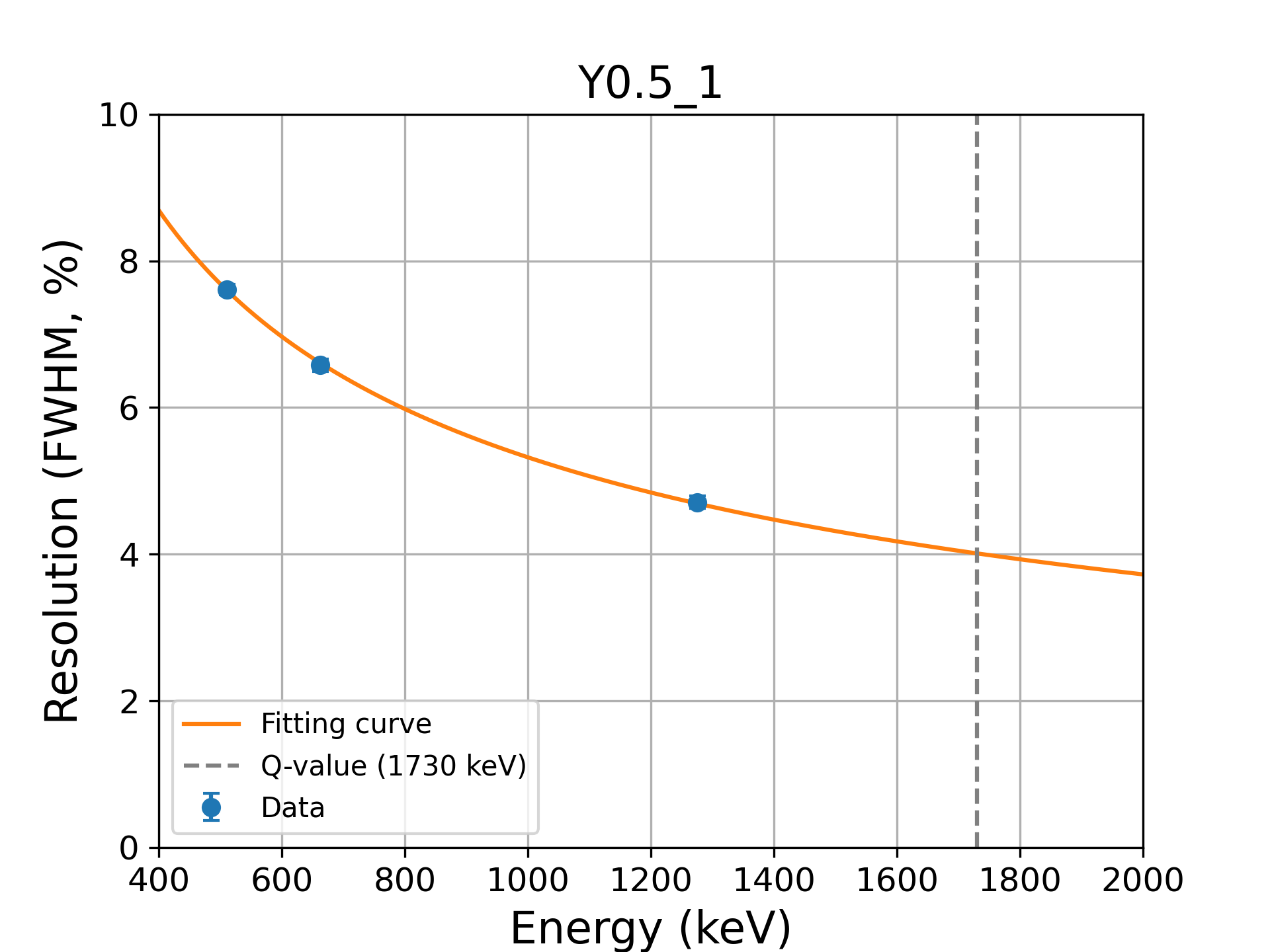}
    \caption{
	Energy dependence of energy resolution.
	%The blue points are the data, and the orange line is the fitting result.
    %The vertical axis represents the FWHM.
	%The gray dashed line indicates the Q-value of $^{160}$Gd (1730 keV).
}
    \label{fig:resolution_curve}
\end{subfigure}
\caption{(a) Example of fitting to obtain energy resolution. The black line is the data, the red line is the fitting function. Energy spectra of $^{137}$Cs. (b) Energy dependence of energy resolution and its fitting. The vertical axis represents the FWHM. The gray dashed line indicates the Q-value of $^{160}$Gd (1730 keV).}
\end{figure}

The energy resolution was evaluated using the photoelectric absorption peaks and defined as the ratio of the full width at half maximum (FWHM) to the peak energy. As demonstrated in the fitting example in Figure \ref{Fitting_example}, the photoelectric peaks were modeled using a Gaussian function combined with a linear background to account for the underlying continuum.
An initial fit was performed with the fitting range determined visually.
The fitting range was then refined to the $\mu \pm 2\sigma$ region obtained from the initial fit, and fitted again.

\begin{figure}
\centering
\includegraphics[width=4in]{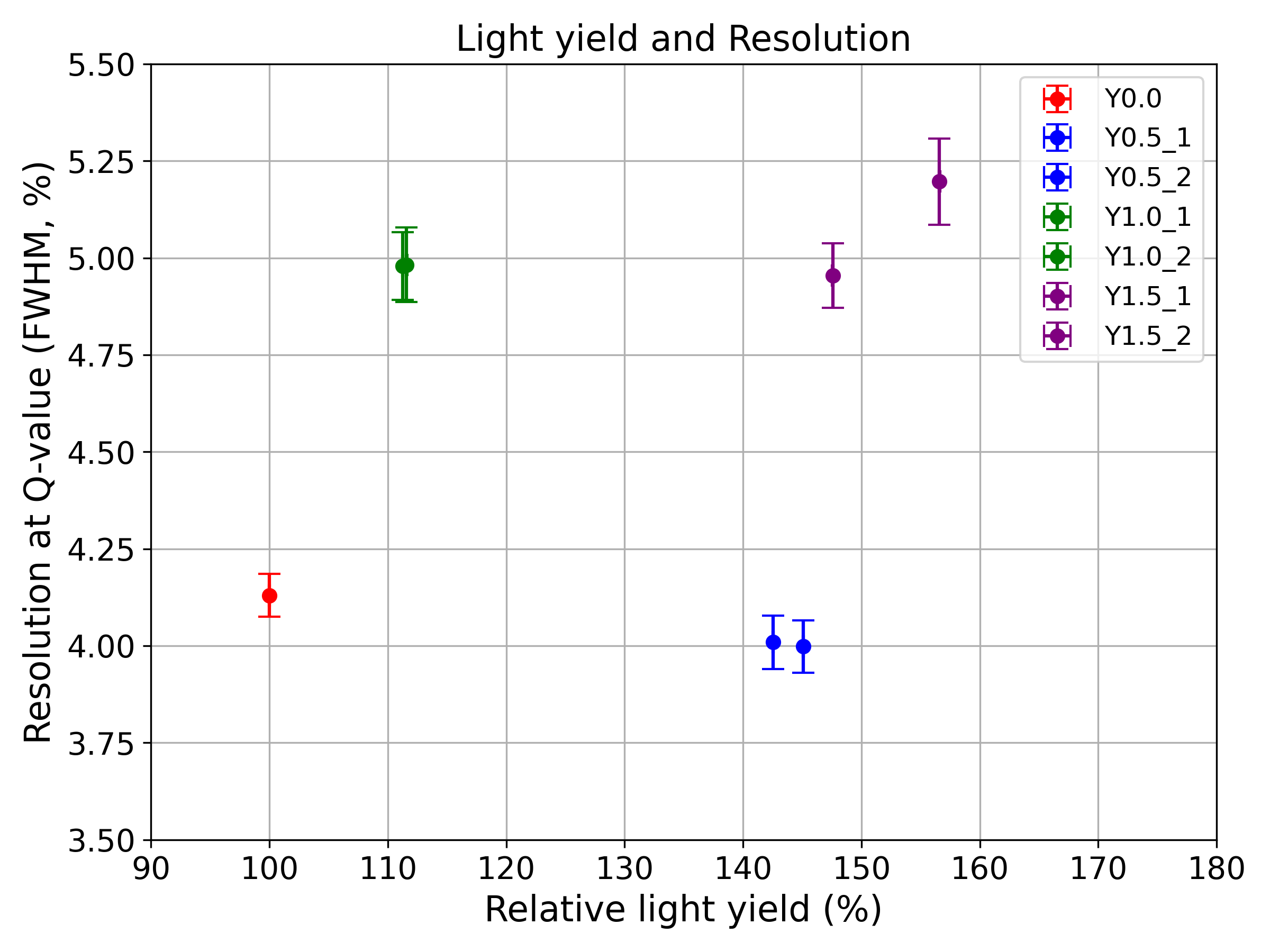}
\caption{Relationship between light yield and energy resolution (FWHM) at the Q-value of $^{160}$Gd.
        The relative light yield represents the relative value of the 662 keV peak position of $^{137}$Cs for each crystal, normalized to GAGG as 100.
        }
\label{fig:light_yield_vs_resolution}
\end{figure}

Energy resolution was measured at three distinct energies: 662 keV from $^{137}$Cs, and 511 keV and 1275 keV from $^{22}$Na.
Figure \ref{fig:resolution_curve} illustrates the relationship between incident energy and energy resolution for the Y0.5 crystal.
To estimate the energy resolution at the Q-value of $^{160}$Gd (1730 keV), the measured data were fitted using the empirical function $\Delta  E / E=a/\sqrt{E}+b$, where $\Delta E$ denotes the FWHM. The resulting estimated energy resolutions for each crystal at the Q-value are summarized in Table 2.
The Y0.5 crystals exhibited the best energy resolution.
Figure \ref{fig:light_yield_vs_resolution} shows the relationship between light yield and energy resolution at the Q-value.
The light yield of Y0.5 is 40\% higher than that of GAGG, and its energy resolution was slightly better than that of GAGG.
While the light yield of Y1.0 and Y1.5 crystals are higher than that of GAGG, their energy resolution is worse than that of GAGG.

\begin{table}[ht]
\caption{Energy resolution and light yield of each crystal.
The light yield is normalized to 100 for the Y0.0 sample.
The errors of the light yields are negligibly small.
}
\label{table_energy_resolution}
\centering
\begin{tabular}{|c||c|c|}
\hline
Crystal & FWHM at Q-value (\%) & Light yield \\ 
\hline
Y0.0 & 4.1 $\pm$ 0.0 & 100\\
Y0.5, 1 & 4.0 $\pm$ 0.0 & 143\\
Y0.5, 2 & 4.0 $\pm$ 0.0 & 145\\
Y1.0, 1 & 5.0 $\pm$ 0.1 & 111\\
Y1.0, 2 & 5.0 $\pm$ 0.1 & 111\\
Y1.5, 1 & 5.2 $\pm$ 0.1 & 156\\
Y1.5, 2 & 5.0 $\pm$ 0.1 & 148\\
\hline
\end{tabular}
\end{table}

%% file: chapters/chapter4.tex
\section{Evaluation of Pulse Shape Discrimination Capability}
\subsection{Method}
\begin{figure}[ht]
    \centering
    \includegraphics[width=0.6\linewidth]{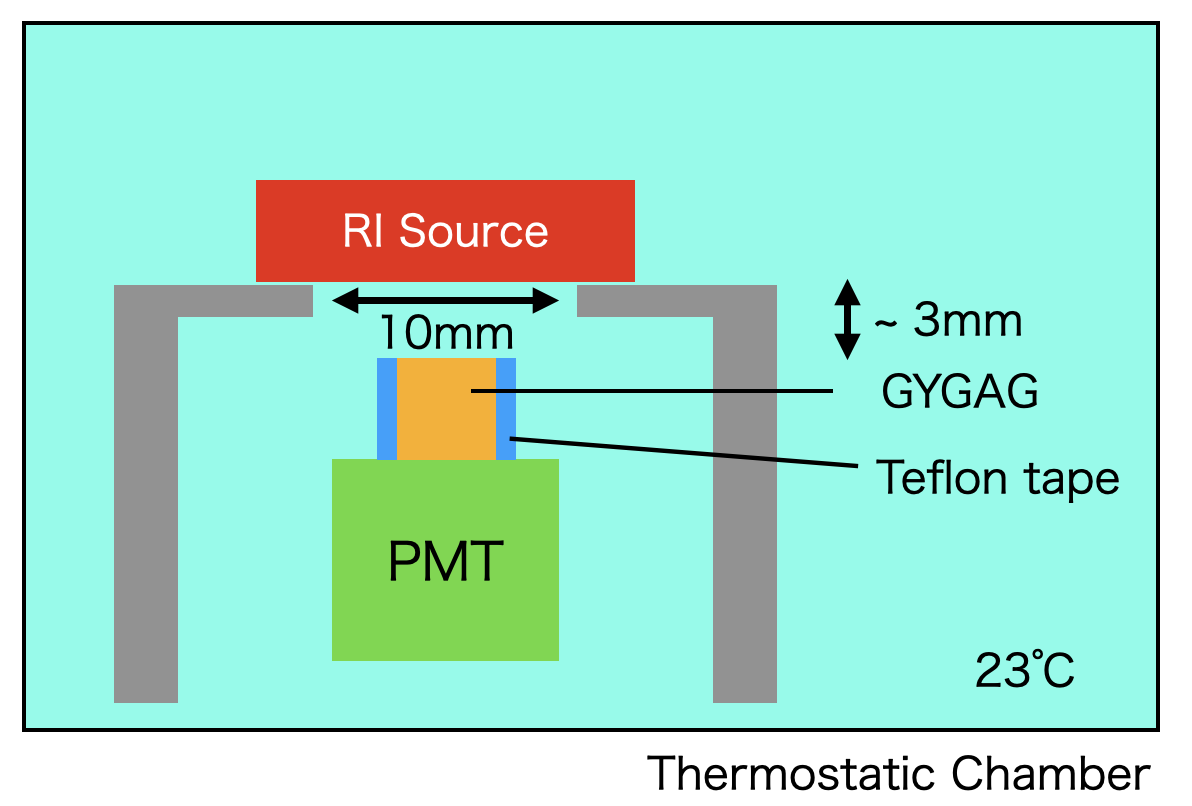}
    \caption{Schematic view of the measurement setup.}
    \label{fig:PMT_setup}
\end{figure}

\begin{figure}[ht]
    \centering
    \includegraphics[width=0.6\linewidth]{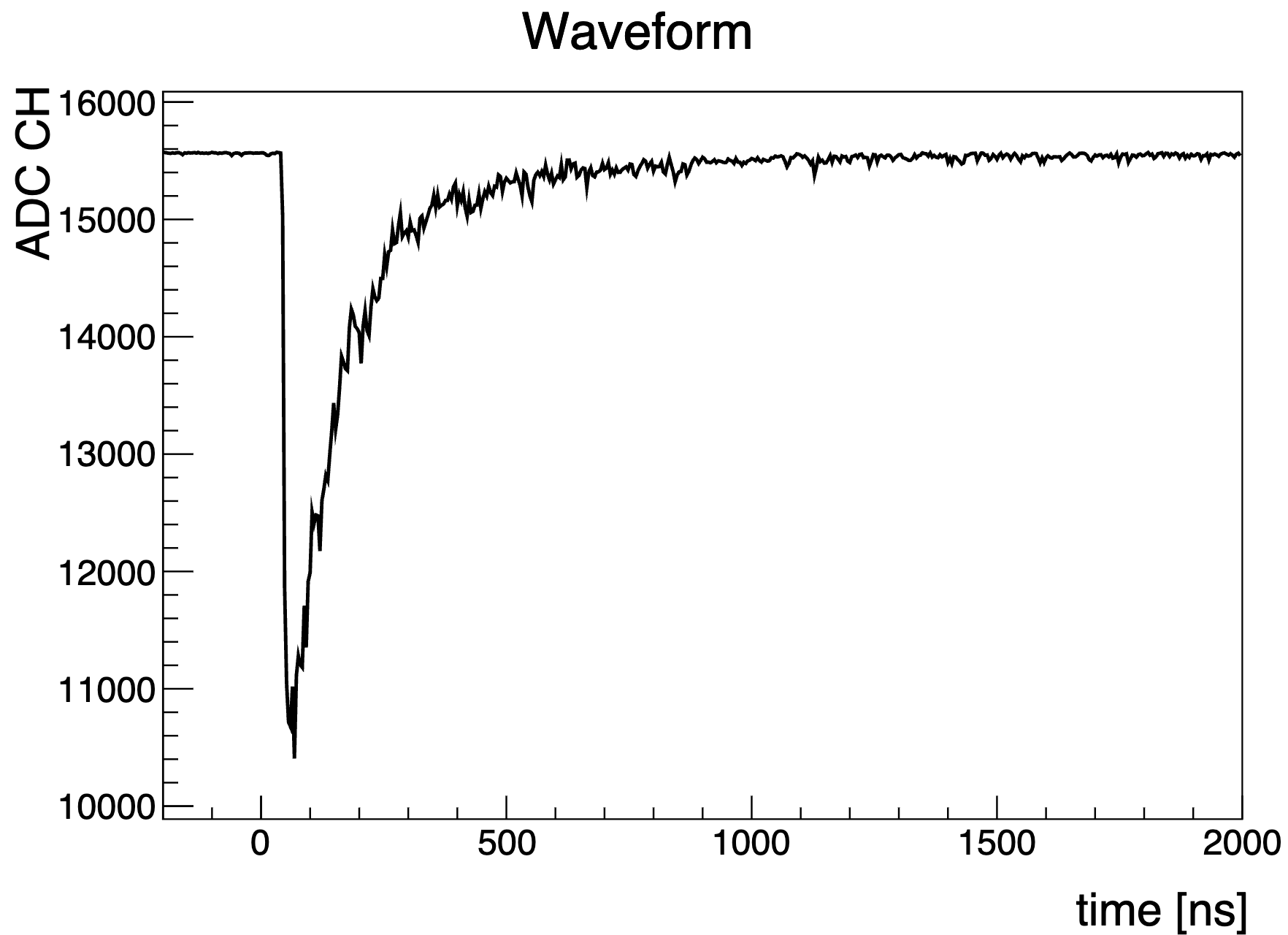}
    \caption{Example of raw waveforms obtained with the $^{241}$Am source.}
    \label{fig:raw_Am_waveform}
\end{figure}

For pulse shape discrimination (PSD) measurements, a photomultiplier tube (PMT; Hamamatsu Photonics H11934-300) was employed due to its high sensitivity at the typical emission wavelengths of GAGG crystals. 
Data acquisition was performed using CAEN digitizers: a DT5725S (250 MS/s) was utilized for both PSD and decay time measurements, while a DT5761 (4 GS/s) was dedicated to rise time measurements.
$^{22}$Na and $^{241}$Am radioactive sources were used for the measurements.
The dimensions of the crystals are the same as those used for the energy resolution measurements.
The distance between the sources and the crystal was fixed at approximately 3 mm.
The activities of the $^{22}$Na and $^{241}$Am sources at the time of measurement were $3.9 \times 10^{5}$~Bq and $1.0 \times 10^{3}$~Bq, respectively.
To optimize light collection efficiency, the crystal was wrapped in Teflon tape; however, the tape on the top surface was removed specifically for alpha-source measurements. The enclosure housing the PMT and the crystal features a square aperture (approximately 1~cm per side) on its top surface, through which the radiation source is introduced. The distance between the source and the crystal was maintained at approximately 3~mm. To ensure stable operation, the PMT assembly was placed in a thermostatic chamber maintained at a constant temperature of 23.0~$\pm$~0.5~$^o$C A schematic of the experimental setup is presented in Figure~\ref{fig:PMT_setup}, and representative raw waveforms obtained with the $^{241}$Am source are shown in Figure~\ref{fig:raw_Am_waveform}.

\subsection{Scintillation Timing Properties} \label{sec:timing_analysis}
\subsubsection{Decay Time Analysis}
\begin{figure}[ht]
    \centering
    \includegraphics[width=0.5\linewidth]{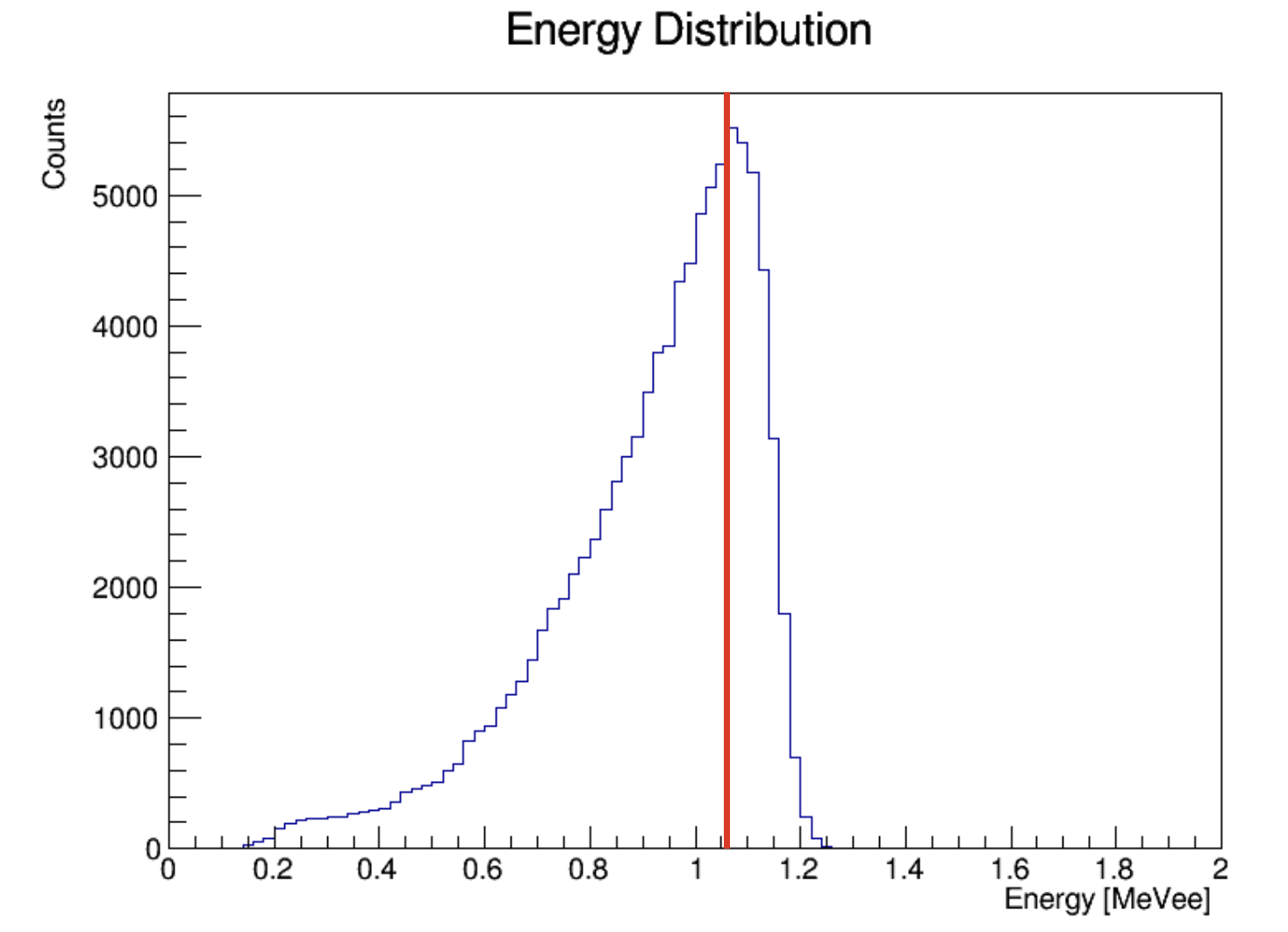}
    \caption{Energy distribution of $^{241}\text{Am}$ measurements. The red line indicates the energy threshold for $\alpha$-rays.}
    \label{alpha_energy}
\end{figure}

The scintillation decay profiles for $\alpha$ and $\gamma$ rays were characterized using average waveforms. 
Figure~\ref{alpha_energy} shows the energy distribution for $^{241}\text{Am}$ measurements after filtering for events above 59.5 keV. Events in the energy region exceeding the peak bin—indicated by the red line in Figure~\ref{alpha_energy}--were identified as alpha particles. Average waveforms were then generated for each crystal using approximately 10,000 events.
To ensure a consistent comparison, $\gamma$ ray events from the $^{22}$Na's Compton scattering region were selected to match the light output of the 5.49~MeV $\alpha$ peak.

To mitigate trigger jitter caused by the digitizer's fixed threshold, the time origin ($t=0$) for each pulse was defined as the timing when the amplitude reached 3\% of its maximum. The resulting average waveforms were then fitted with a double exponential function:
\begin{align}\label{eq:decay_fit}
y(t) = A_1 e^{-\frac{t - t_0}{\tau_1}} + A_2 e^{-\frac{t - t_0}{\tau_2}},
\end{align}
where $A_i$ and $\tau_i$ represent the normalized amplitude ($\sum A_i=1$) and the decay time constant of each component, respectively. The fitting was applied to the decay portion starting from 12~ns ($\gamma$ rays) and 16~ns ($\alpha$ rays) after the pulse maximum, extending up to 1800~ns for both cases to accurately capture the scintillation time profile.

\begin{figure}[ht]
    \centering
    \begin{subfigure}[b]{0.4\textwidth}
        \centering
        \includegraphics[width=\textwidth]{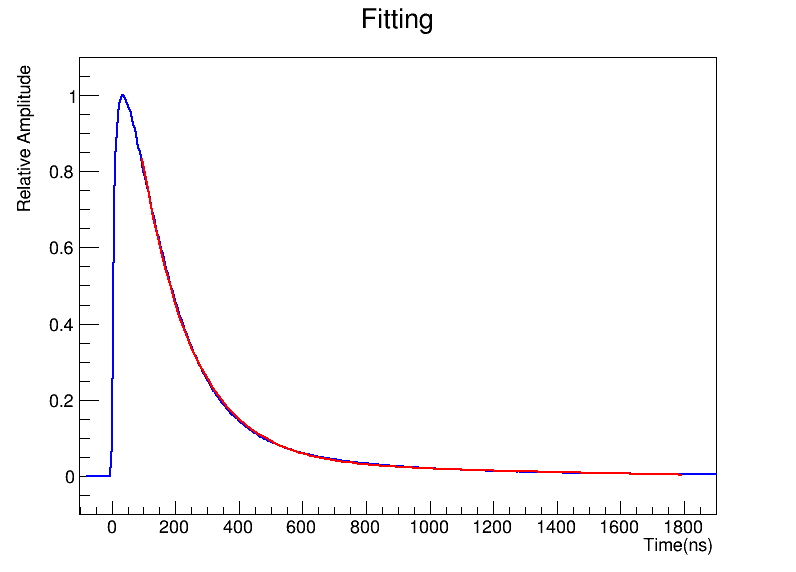}
        \caption{Y0.0, Am $\alpha$-ray source}
        \label{fig:Fitting_example_Am_0.0}
    \end{subfigure}
    \begin{subfigure}[b]{0.4\textwidth}
        \centering
        \includegraphics[width=\textwidth]{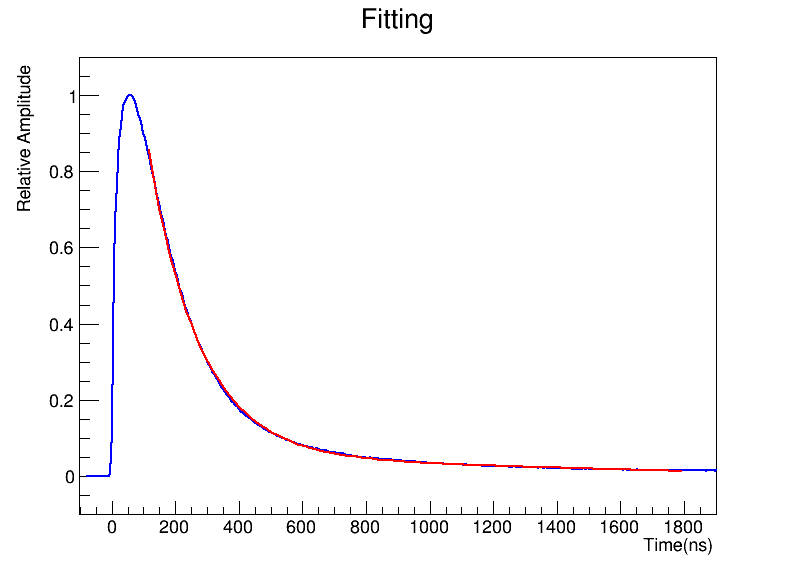}
        \caption{Y0.0, Na $\gamma$-ray source}
        \label{fig:Fitting_example_Na_0.0}
    \end{subfigure}
    \\
    \begin{subfigure}[b]{0.4\textwidth}
        \centering
        \includegraphics[width=\textwidth]{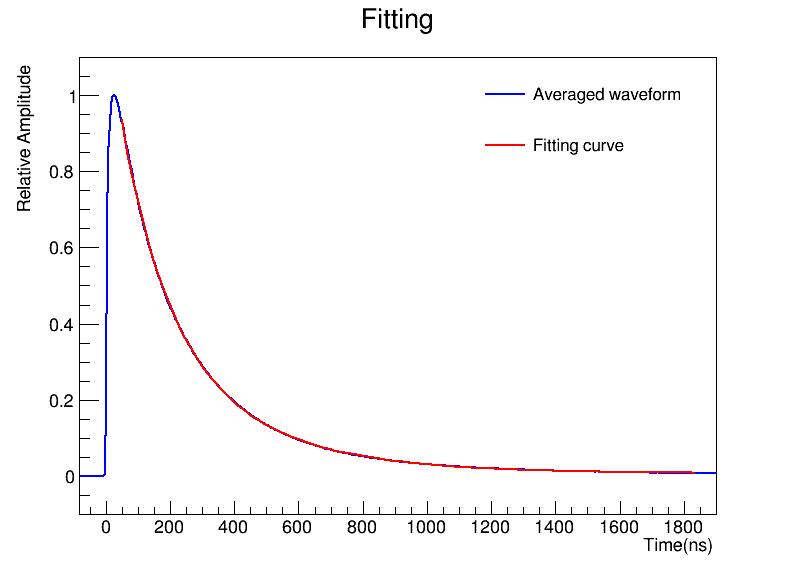}
        \caption{Y0.5, Am $\alpha$-ray source}
        \label{fig:Fitting_example_Am}
    \end{subfigure}
    \begin{subfigure}[b]{0.4\textwidth}
        \centering
        \includegraphics[width=\textwidth]{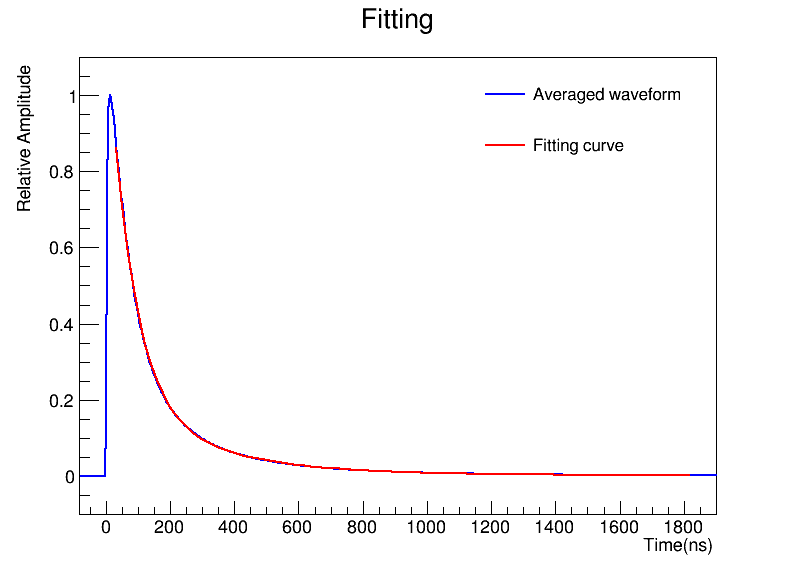}
        \caption{Y0.5, Na $\gamma$-ray source}
        \label{fig:Fitting_example_Na}
    \end{subfigure}
    \caption{Fitting results of decay waveforms. The blue line is the average waveform, and the red line is the fitting result.} 
    \label{fig:waveform_analysis}
\end{figure}

Figure \ref{fig:Fitting_example_Am} and \ref{fig:Fitting_example_Na} show examples of the fitting results for each radiation.
Tables \ref{table_Am_fit} and \ref{table_Na_fit} summarize the fitting results for each crystal.
In GYGAG crystals, differences in the ratio of the two decay components were observed between $\alpha$ and $\gamma$ rays.
For $\alpha$ rays, the proportion of the longer decay component was larger compared to that for $\gamma$ rays, which is also reflected in the average waveforms.
In contrast, in pure GAGG crystals, no significant differences were observed in the component ratios between $\alpha$ and $\gamma$ rays.

\begin{table}[ht]
    \centering
    \caption{Fitting results for each crystal.}
    \begin{subtable}[t]{\linewidth}
        \centering
        \caption{Fitting results of Am source.}
        \label{table_Am_fit}
        \begin{tabular}{|c||c|c|c|c|}
        \hline
        Crystal & amplitude 1 & time constant 1 (ns) & amplitude 2 & time constant 2 (ns)\\ 
        \hline
        Y0.0  & 0.39 &  135.7 $\pm$ 0.4  & 0.61 &  181.1 $\pm$ 0.4 \\
        Y0.5, 1  & 0.34 &  105.6 $\pm$ 0.1   & 0.66 &  265.3 $\pm$ 0.1  \\
        Y0.5, 2  & 0.33 &  96.0 $\pm$ 0.1   & 0.67 &  271.5 $\pm$ 0.1 \\
        Y1.0, 1  & 0.74 &  76.4 $\pm$ 0.0  & 0.26 &  383.8 $\pm$ 0.3 \\
        Y1.0, 2  & 0.76 &  79.0 $\pm$ 0.0  & 0.24 &  404.9 $\pm$ 0.3 \\
        Y1.5, 1  & 0.76 &  79.5 $\pm$ 0.0  & 0.24 &  401.8 $\pm$ 0.2 \\
        Y1.5, 2  & 0.76 &  88.8 $\pm$ 0.1  & 0.24 &  383.3 $\pm$ 0.3 \\
        \hline
        \end{tabular}
    \end{subtable}

    \vspace{8pt}

    \begin{subtable}[t]{\linewidth}
        \centering
        \caption{Fitting results of Na source.}
        \label{table_Na_fit}
        \begin{tabular}{|c||c|c|c|c|}
        \hline
        Crystal & amplitude 1 & time constant 1 (ns) & amplitude 2 & time constant 2 (ns)\\ 
        \hline
        Y0.0   & 0.42 &  116.1 $\pm$ 0.6  & 0.58 &  192.2 $\pm$ 0.6 \\
Y0.5, 1  & 0.75 &  68.6 $\pm$ 0.0  & 0.25 &  246.3 $\pm$ 0.3 \\
Y0.5, 2  & 0.76 &  70.0 $\pm$ 0.0  & 0.24 &  265.4 $\pm$ 0.3 \\
Y1.0, 1  & 0.93 &  61.8 $\pm$ 0.0  & 0.07 &  283.4 $\pm$ 0.2 \\
Y1.0, 2  & 0.88 &  60.0 $\pm$ 0.0  & 0.12 &  178.6 $\pm$ 0.5 \\
Y1.5, 1  & 0.95 &  69.9 $\pm$ 0.0  & 0.05 &  400.5 $\pm$ 0.2 \\
Y1.5, 2  & 0.95 &  66.9 $\pm$ 0.0  & 0.05 &  381.3 $\pm$ 0.3 \\
        \hline
        \end{tabular}
    \end{subtable}
\end{table}

\subsubsection{Rise Time Analysis}
% The rise time of the waveforms was measured using a sampling rate of 4 GS/s, which is different from that used for decay time measurements.
The method of calculating the average waveform for the rise time was the same as that used for decay time measurements.
The fitting function used is expressed as follows:
\begin{align}
    y = A(1-e^{-\frac{t - t_0}{\tau}})
\end{align}
where $A$ and $\tau$ are the amplitude and time constant of the exponential, respectively.
To characterize the rising edge, the fitting interval was set between 10\% of the maximum amplitude and the peak of the waveform.

\begin{figure}
    \centering
    \begin{subfigure}{0.48\textwidth}
        \centering
        \includegraphics[width=\textwidth]{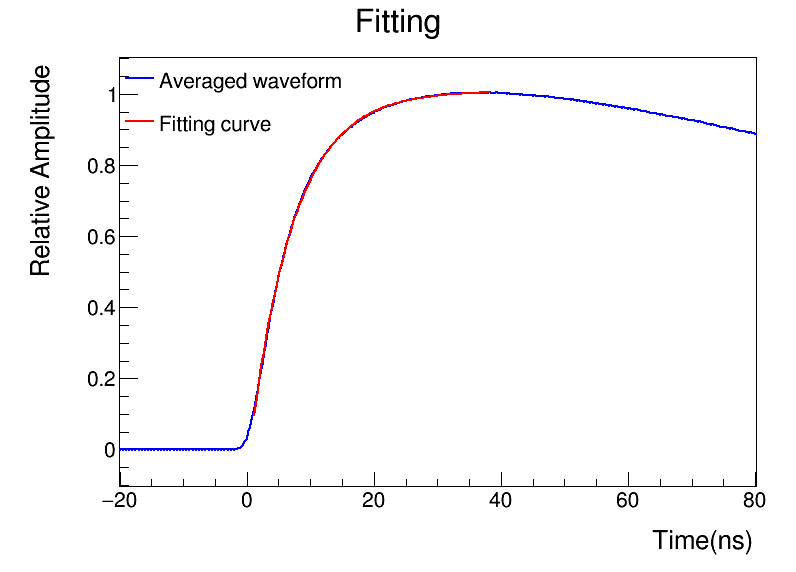}
    \caption{Y0.0, $\alpha$-rays}
    \label{fig:Fitting_example_rise}
    \end{subfigure}
    \hfill
    \begin{subfigure}{0.48\textwidth}
        \centering
        \includegraphics[width=\textwidth]{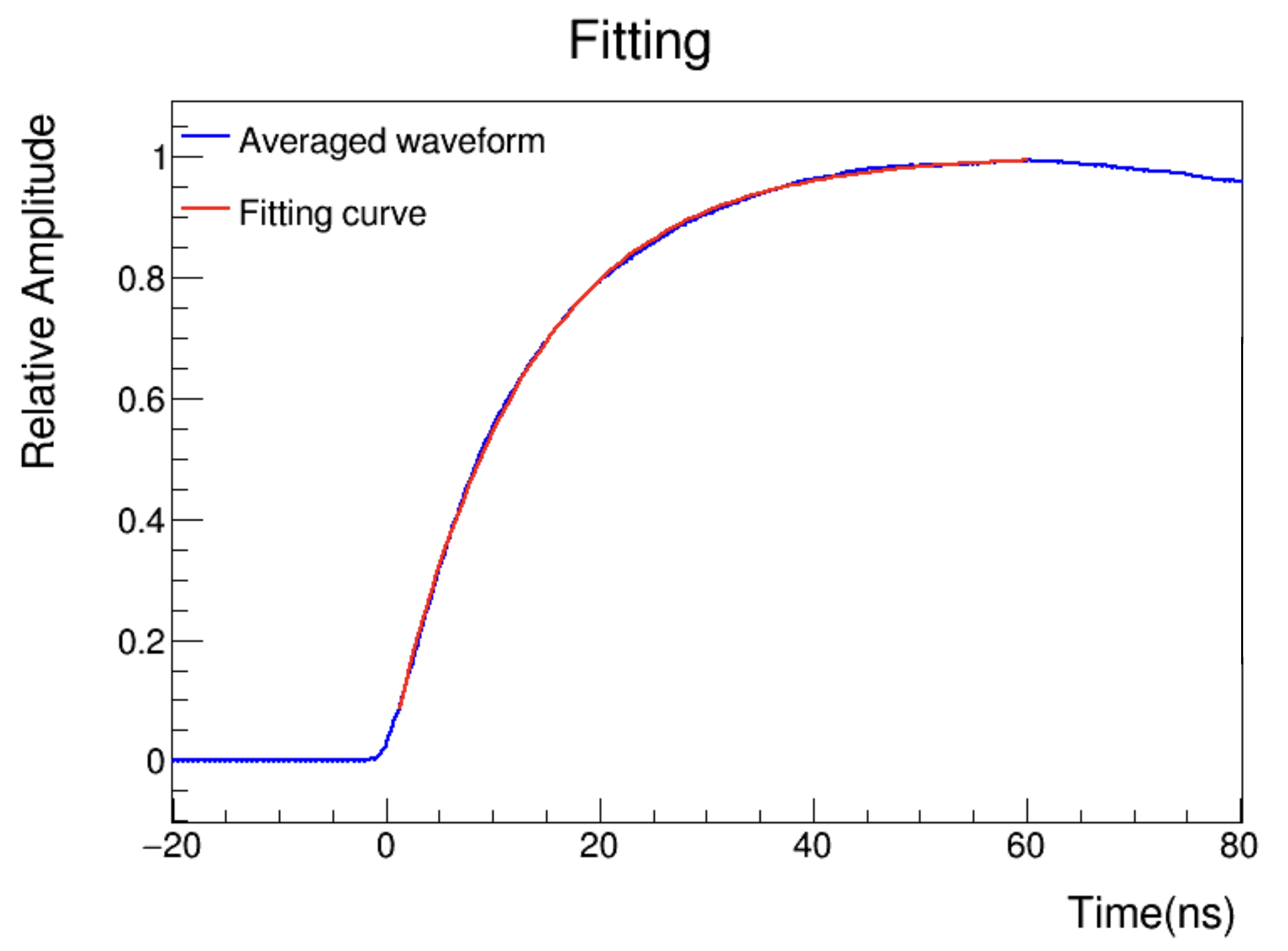}
    \caption{Y0.0, $\gamma$-rays}
    \label{fig:Fitting_example_rise_gamma}
    \end{subfigure}
    \\
    \begin{subfigure}{0.48\textwidth}
        \centering
        \includegraphics[width=\textwidth]{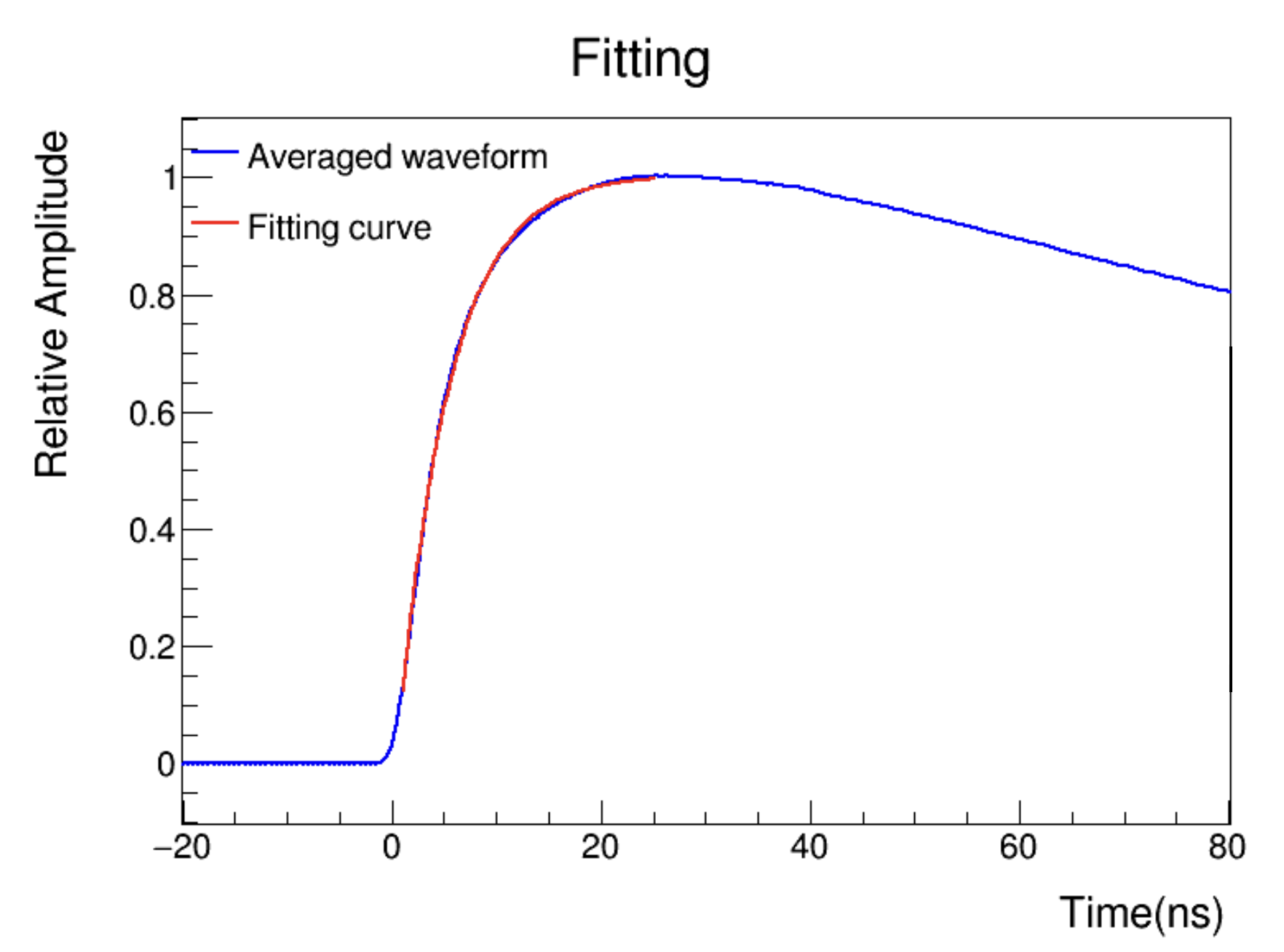}
    \caption{Y0.5, $\alpha$-rays}
    \label{fig:Fitting_example_rise_0.5_alpha}
    \end{subfigure}
    \hfill
    \begin{subfigure}{0.48\textwidth}
        \centering
        \includegraphics[width=\textwidth]{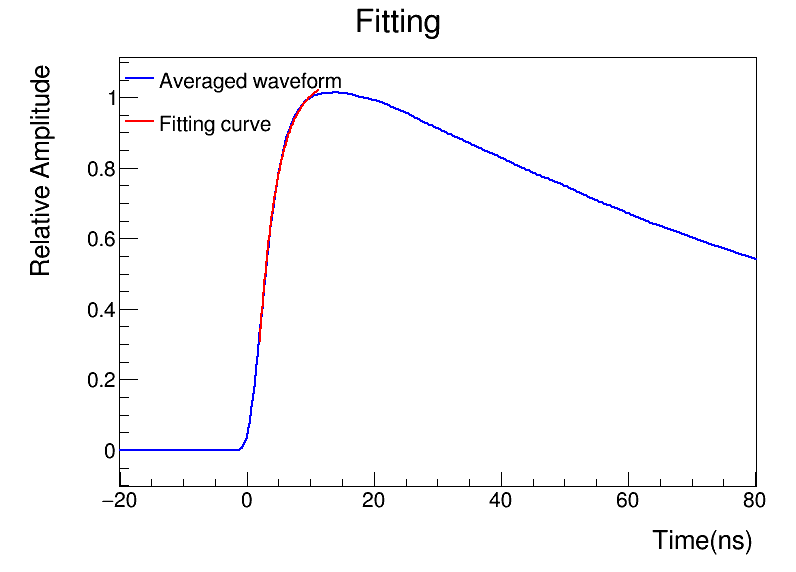}
    \caption{Y0.5, $\gamma$-rays}
    \label{fig:Fitting_example_rise_0.5_gamma}
    \end{subfigure}
    \caption{Fitting result of rise time. (a) and (b) are the average waveform of Y0.0 crystal. (c) and (d) are that of Y0.5, 1 crystal. }
\end{figure}

Figure \ref{fig:Fitting_example_rise} shows an example of the fitting.
Table \ref{table_rise_time_fit} summarizes the fitting results for each crystal.
For pure GAGG crystals, a significant difference in rise time was observed between $\alpha$ and $\gamma$ rays, with the $\gamma$ rays exhibiting a longer rise time.
In contrast, for GYGAG crystals, the rise times for both $\alpha$ and $\gamma$ rays were shorter than those in GAGG, and in most cases, the $\alpha$ rays tended to have longer rise times.
In the Y0.5 crystals, a difference of approximately 60\% was observed between the rise times of $\alpha$ and $\gamma$ rays, whereas little difference was seen in the Y1.0 and Y1.5 crystals.

\begin{table}[ht]
    \caption{Fitting results of rise time for each crystal}
    \label{table_rise_time_fit}
    \centering
    \begin{tabular}{|c||c|c|}
    \hline
    Crystal & time constant ($\gamma$ ray) [ns] & time constant ($\alpha$ ray) [ns]\\ 
    \hline
    Y0.0 & 12.65 $\pm$ 0.03 & 6.85 $\pm$ 0.02 \\
    Y0.5, 1 & 2.77 $\pm$ 0.05 & 4.93 $\pm$ 0.03 \\
    Y0.5, 2 & 2.99 $\pm$ 0.05 & 5.11 $\pm$ 0.03 \\
    Y1.0, 1 & 2.57 $\pm$ 0.08 & 2.49 $\pm$ 0.06 \\
    Y1.0, 2 & 2.59 $\pm$ 0.07 & 2.66 $\pm$ 0.05 \\
    Y1.5, 1 & 2.85 $\pm$ 0.06 & 3.00 $\pm$ 0.03 \\
    Y1.5, 2 & 3.14 $\pm$ 0.03 & 3.11 $\pm$ 0.04 \\
    \hline
    \end{tabular}
\end{table}

\subsection{Pulse Shape Discrimination Analysis}
\label{sec:psd_analysis}

To perform pulse shape discrimination from the acquired waveforms, ratios of integrated charge for different time windows were calculated.
The start time for integration was set to 40 ns before the point where the waveform first exceeds 3\% of its maximum amplitude.
The total area was calculated by integrating the waveform from the start time to 2000 ns after the start time.
The PSD parameter, or the area ratio, was defined as the fraction of the area in the first $t$ ns of the waveform relative to the total area.
The integration time t was scanned to optimize performance for each crystal.

\begin{figure}[ht]
\centering
\begin{subfigure}[b]{0.48\textwidth}
    \centering
    \includegraphics[width=\textwidth]{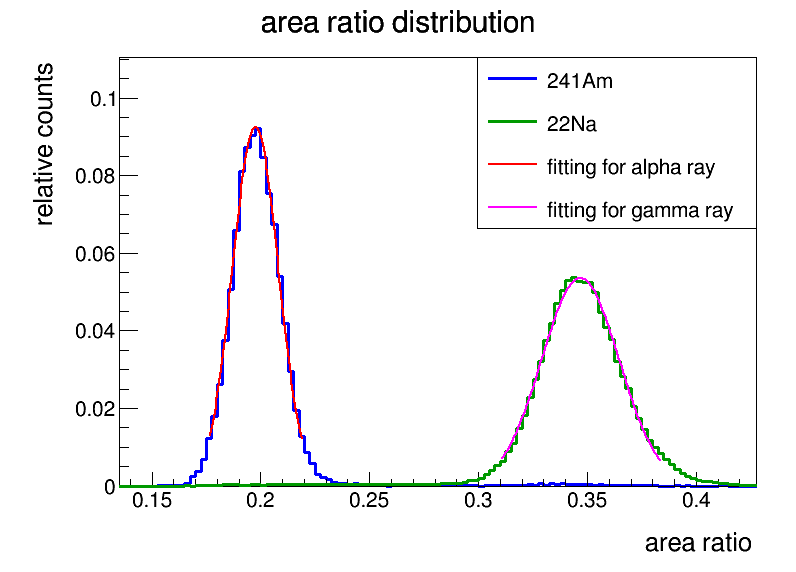}
    \caption{Distribution of the area ratio for Y0.5 crystal with $t = 100$ ns. 
    }
    \label{discri_example}
\end{subfigure}
\hfill
\begin{subfigure}[b]{0.48\textwidth}
    \centering
    \includegraphics[width=\textwidth]{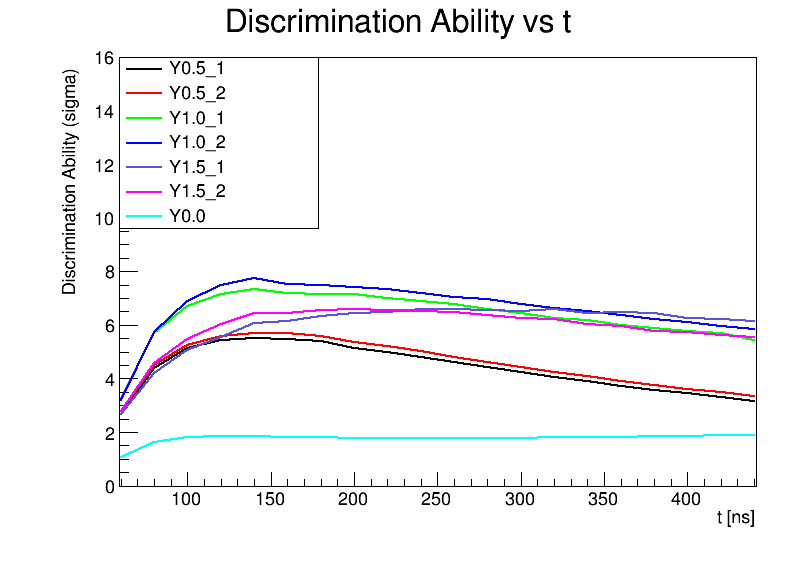}
    \caption{FoM as a function of time window for each crystal. 
    }
    \label{discri_time}
\end{subfigure}
\caption{Pulse shape discrimination analysis: (a) Example of area ratio distribution fitting. The blue histogram is the distribution of $\alpha$ rays, and the green histogram is the distribution of $\gamma$ rays. The red and pink lines are the fitting results using Gaussian functions. (b) The maximum value of the discrimination capability is used as the final value for each crystal.}
\label{fig:psd_analysis}
\end{figure}

Figure \ref{discri_example} shows an example of the area ratio distribution for $\alpha$ and $\gamma$ rays.
The distributions of area ratios for $\alpha$ rays and $\gamma$ rays could be approximated by Gaussian functions.
To evaluate the particle identification performance between $\alpha$ and $\gamma$ rays, we quantified the discrimination capability using the Figure of Merit (FoM). The FoM is defined as:
\begin{equation}\label{eq:discrimination_capability}
\text{FoM} = \frac{|\mu_\alpha - \mu_\gamma|}{\sigma_\alpha + \sigma_\gamma},
\end{equation}
where $\mu_i$ and $\sigma_i$ ($i = \alpha, \gamma$) represent the mean and the standard deviation of the distribution (e.g., the ratio of the light output in different gate widths) for each radiation type, respectively. A higher FoM value indicates a superior separation between the two types of radiation.

Figure~\ref{discri_time} shows FoM as a function of the short integration time width $t[ns]$. For each crystal, the peak value of FoM was adopted as the representative metric of its discrimination performance. These maximum values are summarized in Table~\ref{table_discrimination_capability}. Among the samples, the Y1.0 crystal exhibited the highest performance, followed by the Y1.5 crystal. Notably, even the GYGAG crystal with the lowest FoM value exceeded a score of 5, indicating that these crystals provide more than sufficient separation between $\alpha$ and $\gamma$ rays. In contrast, the GAGG crystal showed a significantly lower discrimination performance compared to the GYGAG series.

\begin{table}
    \caption{Fitting results of Am source for each crystal. The Ga/Al ratio represent the averages of 20 measurement points obtained via EPMA, which were then renormalized so that their sum becomes 5.}
    \label{table_discrimination_capability}
    \centering
    \begin{tabular}{|c|c||c|c|}
    \hline
    Crystal & Ga/Al ratio & Time $t$ (ns) & Max FoM\\ \hline
    Y0.0 & 2.27 / 2.73 & 136 & 1.9\\
    Y0.5, 1 & 2.96 / 2.04 & 144 & 5.5\\
    Y0.5, 2 & 2.94 / 2.06 & 148 & 5.7\\
    Y1.0, 1 & 2.95 / 2.05 & 152 & 7.4\\
    Y1.0, 2 & 2.93 / 2.07 & 152 & 7.8\\
    Y1.5, 1 & 2.96 / 2.04 & 276 & 6.7\\
    Y1.5, 2 & 2.92 / 2.08 & 216 & 6.6\\
    \hline
    \end{tabular}
\end{table}

%% file: chapters/chapter5.tex
\section{Discussions}
The light yield of GYGAG crystals is higher than that of GAGG crystals, and the energy resolution of the Y0.5 crystal is better than that of GAGG crystals.
However, the energy resolutions of Y1.0 and Y1.5 crystals are worse than that of GAGG crystals, despite their higher light yields.
This discrepancy between light yield and energy resolution for the Y1.0 and Y1.5 crystals may be attributed to local compositional non-uniformity. In fact, our EPMA measurements revealed significant segregation of the dopant/compositional elements, with no substantial difference in the average Y concentration between the Y1.0 and Y1.5 samples. Such inhomogeneity likely leads to position-dependent light output within the crystal, resulting in the observed degradation of the overall energy resolution despite the enhanced light yield.

GAGG crystals used in recent double beta decay searches have been reported to exhibit a Pulse Shape Discrimination (PSD) performance with a Figure of Merit (FoM) of 4.6 for the energy between 0.15 and 3 MeV~\cite{Omori2024}. Although direct comparison is difficult, all GYGAG crystals demonstrate FoM exceeding 5 and show a clear dependence on the yittrium concentration. This indicates that GYGAG crystal can exhibit similar or better PSD performance compared to GAGG. While GAGG is generally recognized for its high PSD capability, the FoM of 1.9 observed in our measurement is relatively modest. We attribute this discrepancy to the variation in the Ga/Al ratio. According to prior research~\cite{Yoshino2024}, GAGG with a $\text{Ga}_{2.4}\text{Al}_{2.6}$ composition exhibits significantly suppressed PSD capability—approximately one-fourth of that observed in GAGG with a $\text{Ga}_3\text{Al}_{2}$ ratio. This confirms that our results are consistent with the specific stoichiometry of our sample and underscores the importance of compositional optimization for $0\nu\beta\beta$ decay searches.

%% file: chapters/chapter6.tex
\section{Conclusions}
This study evaluated the scintillation properties of GYGAG crystals with varying Y concentrations. 
Among the tested samples, the Y0.5 crystal—characterized by an EPMA-measured Y concentration of $0.64 \pm 0.02$ —demonstrated the optimal energy resolution. All GYGAG crystals showed PSD performance similar or better than GAGG, and its performance depends on the yttrium concentration.
Although its radioactive impurity is yet to be investigated, the
improved resolution and PSD performance of GYGAG provide a further
reduction in background noise. This makes the GYGAG crystal an
attractive choice for neutrinoless double-beta decay searches for
$^{160}$Gd.

\section*{Acknowledgments}
FE-EPMA measurements were performed at the Department of Earth and Planetary Science, the University of Tokyo. We would like to thank Mr. Koji Ichimura for his support.
This work was supported by the JSPS KAKENHI Grant Number JP21K18625, JP22H01245, and JP23K22516.

%% file: 2603-029-3H-HibikiHayasaki.bbl
\begin{thebibliography}{99}

\bibitem{KamLAND} S. Abe et al. [KamLAND-Zen Collaboration], Phys. Rev. Lett. \textbf{135}, 262501 (2025).


\bibitem{Danevich2001} F. A. Danevich et al., Nucl. Phys. A \textbf{694}, 375 (2001).

\bibitem{GAGG_alpha1} M. Kobayashi et al., Nucl. Instrum. Methods Phys. Res. A \textbf{694}, 91 (2012).

\bibitem{GAGG_alpha2} Y. Tamagawa et al., Nucl. Instrum. Methods Phys. Res. A \textbf{795}, 192 (2015).

\bibitem{Kamioka}
T. Omori et al., Prog. Theor. Exp. Phys. 2024, 033D01 (2024).


\bibitem{GYGAG_1} J.-Y. Zhang et al., J. Eur. Ceram. Soc. \textbf{37}, 4925 (2017).


\bibitem{GAGG_basic1} K. Kamada et al., J. Cryst. Growth \textbf{352}, 88 (2012).

\bibitem{GAGG_basic2} P. Sibczynski et al., Nucl. Instrum. Methods Phys. Res. A \textbf{772}, 112 (2015).
\bibitem{GAGG_alpha3} K. Nakajima et al., Nucl. Instrum. Methods Phys. Res. A \textbf{916}, 51 (2019).

\bibitem{GYGAG_add1}
N. Cherepy et al., IEEE Nucl. Sci. Symp. Conf. Rec. 2010, 1288 (2010).

\bibitem{GYGAG_add2}
G. Hull et al., Proc. SPIE 6706, 67060F (2007). 

\bibitem{Czochralski}
R.~Uecker,
The historical development of the Czochralski method,
J. Cryst. Growth \textbf{401}, 7 (2014).
doi:10.1016/j.jcrysgro.2013.11.095
\bibitem{Kurosawa2014}
S.~Kurosawa, Y.~Shoji, Y.~Yokota, K.~Kamada, V.~I.~Chani, and A.~Yoshikawa,
Czochralski growth of Gd$_3$(Al$_{5-x}$Ga$_x$)O$_{12}$ (GAGG) single crystals and their scintillation properties,
J. Cryst. Growth \textbf{393}, 134 (2014).
doi:10.1016/j.jcrysgro.2013.10.059

\bibitem{ZAF}
R.~Castaing,
Electron Probe Microanalysis,
Adv. Electron. Electron Phys. \textbf{13}, 317 (1960).
doi:10.1016/S0065-2539(08)60212-7







\bibitem{Omori2024}
T. Omori et al., Prog. Theor. Exp. Phys. \textbf{2024}, 033D01 (2024). doi: 10.1093/ptep/ptae026
\bibitem{Yoshino2024}
M.~Yoshino, T.~Iida, K.~Kamada, I.~Shoji, A.~Yoshikawa, and PIKACHU Collaboration, \textit{Meeting Abstracts of the Physical Society of Japan} \textbf{79.2}, 576 (2024).
\end{thebibliography}
